\documentstyle[aps,pra,twocolumn]{revtex}

\begin{document}
\draft
\title{Free expansion of Bose-Einstein condensates with 
quantized vortices}
\author{Franco Dalfovo$^1$ and Michele Modugno$^{2}$}
\address{$^1$Dipartimento di Fisica, Universit\`a di Trento, 
I-38050 Povo, Italy \\ 
and Istituto Nazionale per la Fisica della Materia,
Unit\`a di Trento.}
\address{$^2$Dipartimento di Fisica, Universit\`a di Firenze,
Largo E. Fermi 2, I-50125 Firenze, Italy \\
and Istituto Nazionale per la Fisica della Materia, 
Unit\`a di Firenze. }
\date{ July 6, 1999 }

\maketitle

\begin{abstract}
The expansion of Bose-Einstein condensates with quantized vortices
is studied by solving numerically the time-dependent Gross-Pitaevskii 
equation at zero temperature. For a condensate initially trapped in 
a  spherical harmonic potential, we confirm previous results obtained 
by means of variational methods showing that, after releasing the 
trap, the vortex core expands faster than the radius of the atomic 
cloud. This could make the detection of vortices feasible, by 
observing the depletion of the density along the axis of rotation.
We find that this effect is significantly enhanced in the case of 
anisotropic disc-shaped traps. The results obtained as a function of
the anisotropy of the initial configuration are compared with the
analytic solution for a noninteracting gas in 3D as well as with 
the scaling law predicted for an interacting gas in 2D.  
\end{abstract}
\pacs{03.75.Fi, 03.75.-b , 05.30.Jp }

\section{Introduction}

Since the discovery of Bose-Einstein condensation in trapped gases
of alkali-metal atoms \cite{discovery}, one of the primary goals
of both the experimental and theoretical activity has been the
search for superfluid effects. A clean connection among 
Bose-Einstein condensation and superfluidity would be, in fact, 
of fundamental importance from a conceptual viewpoint, for it 
would improve our understanding of the properties of
quantum many-body systems. In this context, attempts have been
made to produce and detect quantized vortices in these trapped
gases \cite{varenna} and experiments are currently running in
different laboratories.  At the same time, several theoretical 
papers have been written on the expected features of these vortices
(see, for instance, Refs.\cite{fetter,nature,RMP,Stringari} and 
references therein). 

One of the difficulties in observing a vortex in a trapped 
condensate is that its core, i.e., the region where the density is 
depleted by the centrifugal force associated with the quantized
vorticity, has a radius of the order of the healing length $\xi$. 
This quantity is significantly smaller than the radius of the cloud,
especially when the number of atoms, $N$, is large. This makes
a direct observation rather difficult, due to the finite
resolution of the optical imaging devices.

Among the various methods suggested for detecting vortices, one 
consists in letting the condensate expand freely by switching-off the 
trapping potential. It has recently been shown \cite{lundh} that the core
of the vortex may expand faster than the radius of the cloud, making the
observation of the corresponding density depletion feasible. In order
to give quantitative estimates of this effect, in Ref.\cite{lundh}
the Gross-Pitaevskii equation has been solved by using an approximate
trial wave function and the vortex core size has been studied in the
case of a gas initially confined in a spherical trap. 

In the present work we study the same problem, namely the dynamics
of the expanding condensate with a vortex inside, by exactly solving
the Gross-Pitaevskii equation and exploring also the effects of the
anisotropy of the initial configurations. Our calculations for the
spherical case closely agree with the results of Ref.\cite{lundh}, 
showing that the magnification of the core size is not an artifact 
of the variational wave function used by those authors. In the case
of anisotropic configurations, we find different behaviors depending 
on the form of the trap. When the initial confinement is weaker
along the axis of rotation (cigar-shaped traps) the axial expansion
is slower, while the radial one is similar to that of a two-dimensional
condensate: the core size and the radius of the cloud increase with 
almost the same speed, following the scaling behavior expected for 
the solutions of the Gross-Pitaevskii equation in 2D \cite{scaling}.
Conversely, when the initial confinement is stronger along the
axis of rotation (disc-shaped traps), the axial expansion is faster
and the core size is found to increase much more than the radius of 
the cloud in the first instants of motion. The ratio between the
radii of the core and the cloud starts from a rather small value, 
which depends on $N$ and is close to the prediction for a 2D condensate, 
and then increases quickly to the value $0.282$, which is the analytic
result for an ideal gas in 3D. All these results will be discussed
in details in section III, while in the next section we will introduce
the basic formalism and the numerical procedure.

\section{Solving the Gross-Pitaevskii equation}

The dynamics of a dilute and weakly interacting Bose-Einsten 
condensate at zero temperature is well described by a mean-field 
theory, in which the motion of all
the atoms is assumed to be determined by a single complex function
$\Phi ({\bf r},t)$, also called order parameter, playing the role
of a macroscopic wave function and obeying the following equation:
\begin{equation}
i \hbar {\partial \over \partial t} \Phi ({\bf r},t) = 
\left[ - {\hbar^2 \nabla^2 \over 2m } + V_{\rm ext} ({\bf r})
+ g |\Phi ({\bf r},t)|^2 \right] \Phi ({\bf r},t) \; .
\label{eq:GP}
\end{equation}
This is known as Gross-Pitaevskii (GP) equation \cite{GP}. The 
mean-field interaction enters through the term proportional to 
the particle density $n({\bf r},t)=|\Phi ({\bf r},t)|^2$, the
coupling constant $g$ being related to the $s$-wave scattering
length $a$ by $g=4\pi \hbar^2 a /m$. For an axially symmetric
trap the confining potential $V_{\rm ext} ({\bf r})$ can be
written in the form $V_{\rm ext} (\rho, z) = (m/2)[\omega_\rho^2
\rho^2 + \omega_z^2 z^2]$, with $\rho=(x^2+y^2)^{1/2}$. The
ratio between the axial ($\omega_z$) and radial ($\omega_\rho$)
frequencies, $\lambda=\omega_z/\omega_\rho$, is a useful 
parameter fixing the anisotropy of the trap. The derivation and
several applications of the GP equation (\ref{eq:GP}) are reviewed
in Ref.\cite{RMP}.

In general the order parameter entering the GP equation (\ref{eq:GP})
can be written in the form
\begin{equation}
\Phi({\bf r},t) = \sqrt{n({\bf r},t)} \exp [iS({\bf r},t)]
\label{eq:Phi}
\end{equation}
where the phase $S({\bf r},t)$ can be used to define a velocity
field through
\begin{equation}
{\bf v} ({\bf r},t) = { \hbar \over m} \mbox{\boldmath$\nabla$} 
S({\bf r},t) \; .
\label{eq:v}
\end{equation}
When a quantized vortex is present in the condensate, with its
axis of rotation along $z$, the atoms flow with tangential velocity
$v=\kappa \hbar/(m\rho)$, each one having angular momentum $L_z=
\kappa \hbar$, where $\kappa$ is the quantum of circulation. This 
means that, from Eq.~(\ref{eq:v}), the order parameter has to depend 
on the angle $\phi$, around the $z$-axis, in a simple way:
\begin{equation}
\Phi({\bf r},t) = \Psi (\rho,z,t) \exp [i \kappa \phi] \; .
\label{eq:Psi}
\end{equation} 
Inserting this expression back into the GP equation (\ref{eq:GP}),
one gets 
\begin{eqnarray}
i \hbar {\partial \over \partial t} &\Psi & (\rho,z,t) =  
\left[ - {\hbar^2 \nabla^2 \over 2m } + {\kappa^2 \hbar^2
\over 2 m \rho^2 } \right. 
\nonumber \\
& & \left.  + {m\over2} (\omega_\rho^2 \rho^2 + \omega_z^2 z^2)
+ g |\Psi(\rho,z,t)|^2 \right] \Psi(\rho,z,t) \; ,
\label{eq:GP-Psi}
\end{eqnarray}
which corresponds to the GP equation with a centrifugal force 
included. Due this centrifugal term, the solution 
$\Psi(\rho,z,t)$ must vanish on the $z$-axis. The region where the
density is depleted, close to the vortex line, is named vortex
core. In a stationary state, its radius is of the order of
the healing length $\xi$ fixed by the balance between the mean-field
energy and the kinetic energy associated with the gradient of the
density (quantum pressure). In a uniform gas, the healing length
is given by $\xi=(8\pi n a)^{-1/2}$, where $n$ is the density.  
One can use the same expression in order to estimate
the core size in trapped gases by taking for $n$ the central density 
of the condensate without vortices. With  the parameters of the 
current experiments the predicted core radius turns out to be 
very small indeed, making the direct observation of a vortex
rather difficult with the presently available optical 
devices. However, when the condensate is let to expand by 
switching-off the confining potential, the vortex core expands
as well. We want to describe this process by numerically solving
Eq.~(\ref{eq:GP-Psi}).

The GP equation (\ref{eq:GP-Psi}) can be formally rewritten in this
way:
\begin{equation}
i \hbar {\partial \Psi \over \partial t } = H \Psi
\label{eq:NLSE}
\end{equation}
where the effective Hamiltonian $H$ is the operator enclosed in 
square brackets in (\ref{eq:GP-Psi}). Equation (\ref{eq:NLSE}) has
the form of a nonlinear Schr\"odinger equation, since $H$ contains
the unknown $\Psi$. There exist several techniques to solve it 
numerically. One of them consists in propagating the wave function
in time using small time steps $\Delta t$ in order to approximate
the time evolution of $\Psi$ with the equation
\begin{equation}
\left( 1+ {i \Delta t \over 2} H \right) \Psi_{n+1} = 
\left( 1- {i \Delta t \over 2} H \right) \Psi_n 
\label{eq:steps}
\end{equation}
where $H$ and $\Psi_n$ are the effective Hamiltonian and the order 
parameter, respectively, both calculated at the $n$-iteration.
Actually, since the diffusion of $\Psi$ occurs in both the axial 
and radial directions, one splits each time interval in two parts, 
one for the evolution along $z$ and the other along $\rho$. In order 
to propagate $\Psi$ in each direction, we separate the effective 
Hamiltonian into axial and radial components, equally dividing the
mean-field potential (this separation is somewhat arbitrary and 
does not affect the final results).
Equation (\ref{eq:steps}), for both the axial and radial motion at 
each time step, can be solved by mapping the order parameter
on a two-dimensional grid of points, $N_\rho \times N_z$, in
such a way that its solution becomes equivalent to a matrix
diagonalization. This is known as 
Crank-Nicholson differencing method with alternating direction
implicit algorithm \cite{recipes} and was used for the 
expansion of a trapped condensate first by Holland and Cooper
\cite{Holland1}. 

For a vortex in a trap with radial frequency $\omega_\rho$, 
the relevant length scale is $a_\rho=[\hbar/(m\omega_\rho)]^{1/2}$. 
Available condensates have radii of the order of 1 to 10 
$a_\rho$, depending on the strength of the mean-field interaction
fixed by the parameter $Na/a_\rho$ \cite{RMP}.
We consider condensates with repulsive interaction ($a>0$)
and with $Na/a_\rho$ from $0$ to $20$. We choose a grid having
typically $100 \times 200$ points with spacing smaller than
$0.2 a_\rho$, which is sufficient for an accurate description
of the density profile of the condensate including the 
vortex core.  The derivatives in the effective Hamiltonian $H$
are expressed through finite difference 5-points formulas
\cite{abramovitz}. 

The initial configuration at $t=0$ corresponds to the stationary
order parameter of the condensate in the trap for a given $\kappa$.
It can be obtained by using the same algorithm, but diffusing the 
function $\Psi$ in imaginary time instead of real time ($t \to 
- i \tau$). One starts from a reasonable trial wave function and let 
$\Psi$ converge to the stationary state. The same can be done
by using an explicit differencing algorithm instead of the
implicit one based on Eq.~(\ref{eq:steps}). The explicit method
has been used, for instance, in Ref.\cite{df} and turns out 
to be much faster. We used both methods as a test of the 
numerical code. 

For $t>0$ the confining potential in Eq.~(\ref{eq:GP-Psi}) is 
removed and hence the function $\Psi$ expands freely, subject only to 
the mean-field repulsion and quantum pressure. A CPU time of the 
order of $2$ hours on a workstation is typically needed to follow 
the expansion of the condensate for a time of the order of 
$10 \omega_\rho^{-1}$, which is enough to see the effects that
we are going to discuss.

The expansion of an ideal gas ($a=0$) is used as one of the tests
we made on the numerics. In this case, one has simple analytic
results. The $t=0$ configurations are the eigenstates of the 
harmonic oscillator in axial symmetry. The $\kappa=0$ case is a 
Gaussian, while the vortex state $\kappa=1$ corresponds to 
the eigenfunction with azimuthal angular momentum $m=1$. When the
trap is released, their expansion is just the dispersion of
free wave packets. We checked that our numerical integration gives
results practically indistinguishable from the analytic ones. 
In order to verify the proper inclusion of the mean-field
interaction in the code, we made the calculation for the expansion
of a condensate without vortices, using the same parameters
of a previous work \cite{Holland2} and finding exactly the same
results. As a further check, we looked at the behavior of the
energy of the condensate as a function of time. It is well known
that the Crank-Nicholson method may suffer from a possible 
amplification of random numerical noise, due to nonlinearity,
leading to a spurious input of kinetic energy and limiting the
stability of the algorithm to short time. However, we checked 
that this time is long enough to ensure the stability of the 
solutions in the time intervals here considered.

\section{Results}

First we give two examples of expanding vortices. In 
Fig.~\ref{fig:contour1} and \ref{fig:contour2} we show a sequence 
of contour plots of the density $n(\rho,z,t)=|\Psi(\rho,z,t)|^2$
in three instants of time ($t=0,\, 1.5,\, 3 \, \omega_\rho^{-1}$) and 
for two different condensates ($\lambda=0.2$ and $2.5$, respectively). 
The darkest color represents the region where the density is larger 
than $1/2$ of the peak density at $t=0$, and each contour line 
is drawn where the density falls by a factor $2$. Both condensates
have $Na/a_\rho=20$, so that the effects of the mean-field interaction
on the vortex structure are of the same order. The empty region 
close to $\rho=0$ is the vortex core, whose radius grows when the
atomic cloud expands.   The condensate in Fig.~\ref{fig:contour1} 
is initially confined in a cigar-shaped trap. When the trap is released, 
its motion is mainly in the radial direction, the axial size increasing 
much more slowly \cite{note2}. The opposite happens for the condensate
in Fig.~\ref{fig:contour2}: the initial configuration is oblate
(disc-shaped) and the axial motion is faster. 

The shape of the condensate is better seen by plotting
the {\sl column density}, that is the density integrated along 
the $z$-direction, $n_{\rm col}(\rho,t) =\int dz \, n(\rho,z,t)$. 
This quantity is directly measured in the experiments. 
In Fig.~\ref{fig:column}, we plot $n_{\rm col}$  for the same  
expanding condensate shown in  Fig.~\ref{fig:contour2}. 
To give an idea of the actual size, let us suppose to have 
a trap with frequency $\omega_\rho=2\pi (20$Hz$)$ and 
condensates of $^{87}$Rb or Na atoms; then the length scale
$a_\rho$ is $2.4\, \mu$m and $4.7\, \mu$m, respectively. In 
the same cases, the parameters $Na/a_\rho=20$ and $\lambda=2.5$, 
as in Fig.~\ref{fig:column}, would correspond to $N \simeq 8000$ 
for $^{87}$Rb and $N \simeq 34000$ for Na.

By using the column density one can give  suitable definitions of 
the radii of both the atomic cloud and the vortex core. The former
can be identified with the root mean square radius, 
$\rho_{\rm rms}$, obtained by integrating $\rho^2$ over the 
column density distribution, while the core radius, $\rho_c$, can 
be chosen as the value of $\rho$ where the column density first
reaches  $e^{-1}$ 
times the peak value.  It is then useful to study the evolution 
in time of the ratio $\rho_c/\rho_{\rm rms}$, in order to have an 
estimate of the size of the ``hole" made by a vortex in typical 
condensates of current experiments, as done in 
Ref.~\cite{lundh}. In Fig.~\ref{fig:ratio} we show
this ratio for different values of $\lambda$ and $Na/a_\rho$.
This figure contains the main results of the present work.

First, the horizontal dot-dashed line is the analytic 
result for the free dispersion of a wave packet of noninteracting
particles. In this case, the motion in the two directions is decoupled 
and the ratio $\rho_c/\rho_{\rm rms}$ is approximately $0.282$
independently of $\lambda$. The points close to this value 
correspond to the numerical solution of the GP equation (\ref{eq:GP-Psi})
without the mean-field term ($g=0$). The fluctuations at short time 
are not due to random noise in the evolution of $\Psi$, which remains
indeed very close to the analytic solution at any time,  but to the 
smallness of the vortex core at the beginning of the expansion: 
both the position of the peak and that of $\rho_c$ are 
located in between grid points and the use
of interpolation formulas yields small errors. As one can see from
the figure, these fluctuations are rather small and do not affect the
main results of this work.

The results obtained by expanding condensates with $Na/a_\rho=5$ and
$20$ are also shown. At $t=0$, each sequence of points starts from 
the value of $\rho_c/\rho_{\rm rms}$ predicted for a stationary state in
the trap. This value is a decreasing function of both the parameters
$Na/a_\rho$ and $\lambda$. An estimate of this dependence can be 
obtained in the so-called Thomas-Fermi limit $Na/a_\rho\gg 1$, when 
quantum pressure is negligible compared to the mean-field repulsion.
In this case, one finds $\rho_c /\rho_{\rm rms} \sim \xi/\rho_{\rm rms} 
\propto (15 \lambda N a/a_\rho )^{-2/5}$. Thus the portion of 
condensate where the density is depleted by the vortex becomes 
smaller and smaller when $N$ and/or $\lambda$ increase. 
This is why in current experiments the vortex core  might be hardly 
observable by simply looking at the density distribution in the
trap.

At $t>0$ the ratio $\rho_c/\rho_{\rm rms}$ increases in different ways
depending on the anisotropy of the trap. In the case of a spherical 
trap ($\lambda=1$), we compare the numerical solution of the GP equation 
(points) with  the results of the variational calculation by Lundh, 
Pethick and Smith \cite{lundh} (solid lines). The agreement is remarkable.
The ratio $\rho_c/\rho_{\rm rms}$ is shown to increase rather quickly 
towards the analytic result for the noninteracting gas.  As discussed 
in Ref.~\cite{lundh}, this is due to the fact that, in the first instants 
of motion, the core size adjust almost istantaneously to the value
$\xi  = (8\pi n a)^{-1/2}$, with $n$ given by the local density just
outside the vortex core. This density decreases rapidly, yielding an 
increase of $\rho_c$ faster than the one of $\rho_{\rm rms}$. After
a characteristic decoupling time, this mean-field effect tends to vanish,
since the gas becomes more and more dilute, and the condensate expands as
an ideal gas, with the axial and radial motions decoupled. The ratio 
$\rho_c/\rho_{\rm rms}$ thus remains almost frozen to its value at the 
decoupling time. 

As one can see in Fig.~\ref{fig:ratio}, the magnification of the core 
size depends on both $Na/a_\rho$ and $\lambda$. In order to 
emphasize the dependence on the anisotropy of the trap, in 
Fig.~\ref{fig:3lambda} the results for $Na/a_\rho=20$ and different
values of $\lambda$ are plotted together. The lowest dashed line is the
result we obtain by solving the GP equation in a purely 2D case with the
same $Na/a_\rho$. In this case, one can prove that the dynamics is
governed by an exact scaling law \cite{scaling}, that is, the 
condensate preserves its shape with a time dilatation of length scales; 
thus $\rho_c/\rho_{\rm rms}$ remains constant. Our results show that 
the ratio $\rho_c/\rho_{\rm rms}$ is almost constant even for condensates 
initially confined in prolate (cigar-shaped) traps, the value at $t=0$ 
being fixed by the local density near the vortex in the stationary state. 
This can be understood by noticing that,  for $\lambda \ll 1$, the axial 
motion is so slow that it does not affect significantly the behavior of 
the radial expansion. In other words, the lowering of the local density, 
which causes $\rho_c$ to increase, is almost completely determined by 
the radial expansion; the condensate can be thought as made by thin 
slices, orthogonal to the $z$ axis, each one expanding as a 2D system.
The opposite  happens in the case of disc-shaped
traps ($\lambda \gg 1$), where the initial value of $\rho_c/\rho_{\rm rms}$
is indeed close to the one of a 2D system, but quickly approaches the
3D ideal gas prediction. This is again a consequence of the axial 
motion. In this case in fact, the rapid decrease of the local density 
around the vortex is almost entirely due to the fast axial expansion
and hence the core size increases much faster than in a 2D system. 

The fact that the 2D scaling behavior applies to the expansion of
cigar-shaped condensates rather than the disc-shaped ones was not 
obvious so far.  For instance, in a recent paper by Castin and Dum 
\cite{castin}, focused
on the properties of vortices in disc-shaped traps, the expansion
of the condensate has been treated by keeping the confinement along
$z$ constant and using the 2D scaling law for the radial motion. 
However, as shown in the present work, the motion along $z$ in the
actual 3D experiments is not expected to be a small correction to 
the 2D scaling law. On the contrary, it is expected to dominate
the dynamics of the expansion, yielding values of $\rho_c/\rho_{\rm rms}$
close to $0.282$ in short time and almost independently of $N$.
This may help the direct detection of quantized vortices in these
inhomogeneous Bose-Einstein condensates, the size of the ``hole"
in the density distribution becoming easily larger than the spatial
resolution of the imaging devices. In order to show how this 
expanding holes might appear in an experimental observation, the
column density of Fig.~\ref{fig:column} is plotted again in 
Fig.~\ref{fig:3D} as a 3D plot, the three frames corresponding to
$t=0, \, 1.5$ and $3 \, \omega_\rho^{-1}$.

\acknowledgements
This work has been supported by Istituto Nazionale per la
Fisica della Materia through the Advanced Research Project
on BEC, and by Ministero dell'Universit\`a e della Ricerca
Scientifica e Tecnologica. F.D. would like to thank the
Aspen Center for Physics where this work has been completed.

\begin{figure}
\caption{Contour plot of the density  $n(\rho,z,t)=|\Psi(\rho,z,t)|^2$
in three instants of time (from left to right, $t=0,\, 1.5,\, 3 \, 
\omega_\rho^{-1}$) for a condensate with $\lambda=\omega_z/\omega_\rho=0.2$ 
and $Na/a_\rho=20$ and having a quantized vortex ($\kappa=1$) along 
the $z$-axis. The darkest color represents the region where the density 
is larger than $1/2$ of the peak density at $t=0$, and each contour line 
is drawn where the density falls by a factor $2$. The co-ordinates
$z$ and $\rho$ are given in units of $a_\rho=[\hbar/
(m\omega_\rho)]^{1/2}$, where $\omega_\rho$ is the radial trapping 
frequency at $t=0$.  }
\label{fig:contour1}
\end{figure}

\begin{figure}
\caption{Same as in Fig.~\protect\ref{fig:contour1} but for a condensate 
in a trap with $\lambda=\omega_z/\omega_\rho=2.5$.  }
\label{fig:contour2}
\end{figure}

\begin{figure}
\caption{Column density $n(\rho,t)$ obtained by integrating along $z$
the density plotted in Fig.~\protect\ref{fig:contour2}. The three curves
correspond to  $t=0,\, 1.5,\, 3 \, \omega_\rho^{-1}$.   }
\label{fig:column}
\end{figure}

\begin{figure}
\caption{ Ratio between the vortex core radius and the root mean square
radius of the condensate, $\rho_c/\rho_{\rm rms}$, as a function of time,
for three different initial configurations ($\lambda=0.2,\, 1,\, 2.5$) and
for $Na/a_\rho= 0, \, 5,$ and $20$. Time is in units of $\omega_\rho^{-1}$. 
Points are the numerical results of the present work. The horizontal 
dot-dashed line corresponds to the expansion of an ideal gas in the 
state $m=1$. The two solid lines for $Na/a_\rho= 5$ and $20$ in the 
spherical case, $\lambda=1$, are the results of recent
variational calculations \protect\cite{lundh}.  }
\label{fig:ratio}
\end{figure}

\begin{figure}
\caption{ Ratio $\rho_c/\rho_{\rm rms}$ as a function of time, for three 
different initial configurations  ($\lambda=0.2,\, 1,\, 2.5$) but the
same value of $Na/a_\rho=20$.  The dot-dashed line is the analytic result
for an expanding ideal gas. The dashed line is the
result obtained by solving the GP equation in a purely 2D system.  }
\label{fig:3lambda}
\end{figure}

\begin{figure}
\caption{ The column density of Fig.~\protect\ref{fig:column} is  
shown here as a 3D plot in order to emphasize the structure of the
expanding vortex. The three frames correspond again to $t=0, \, 1.5$ 
and $3 \, \omega_\rho^{-1}$. }
\label{fig:3D}
\end{figure}


\begin{references}
\bibitem{note2} Note that the apparent contraction of the condensate
along $z$ in Fig.~\ref{fig:contour1} is an artifact of the contour plot.
In fact, the density is decreasing during the expansion and thus the 
area within each contour line tends to decrease as well. The contour 
plot is suitable for representing the behavior of the aspect ratio 
(axial to radial size), but not for the absolute average size in each 
direction.
\end{references}
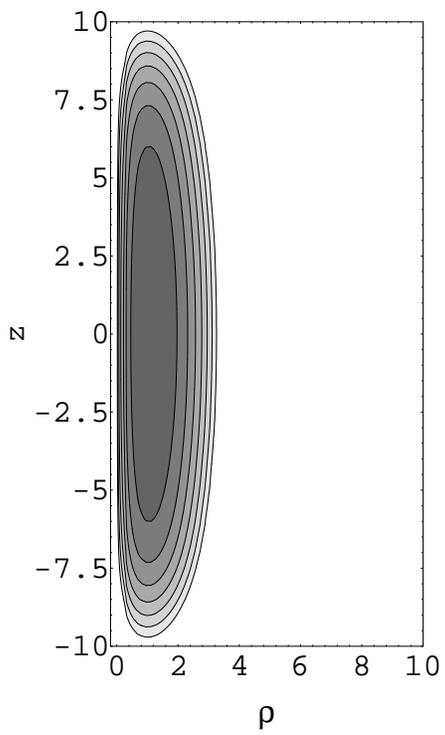
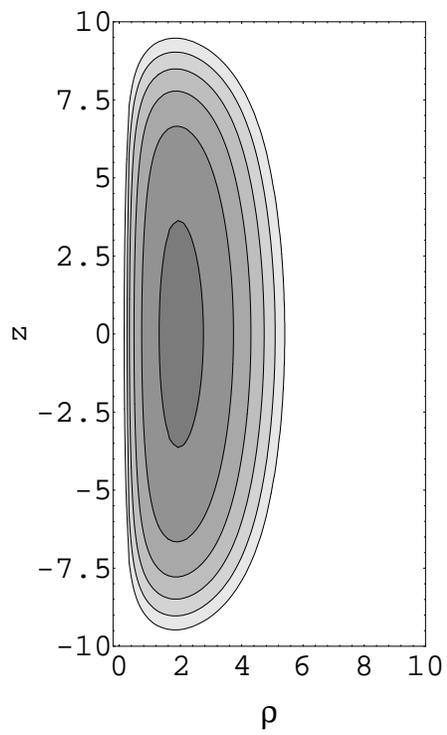
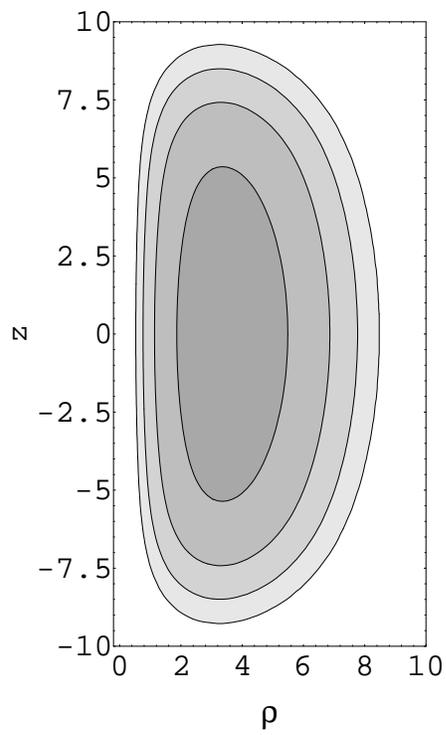
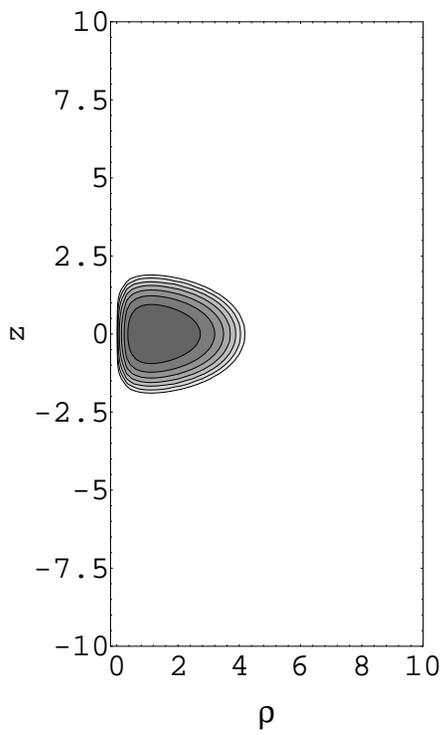
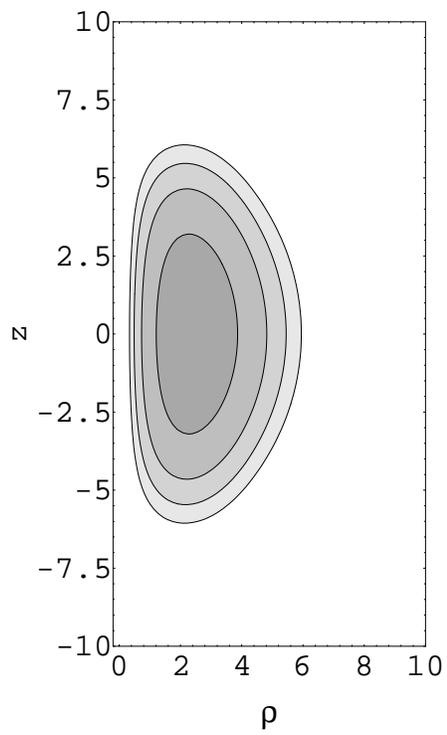
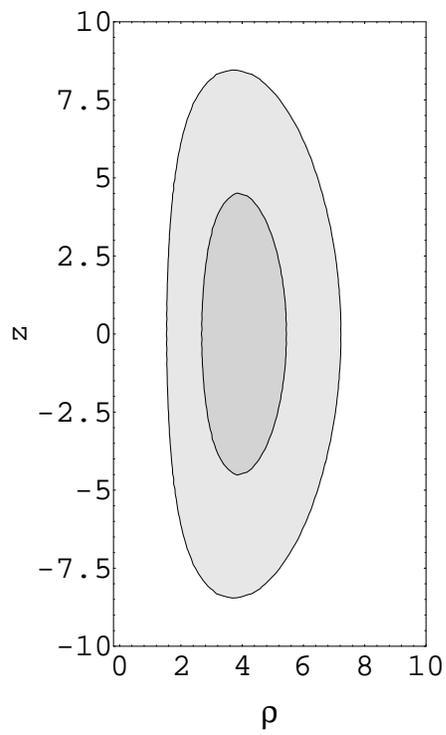
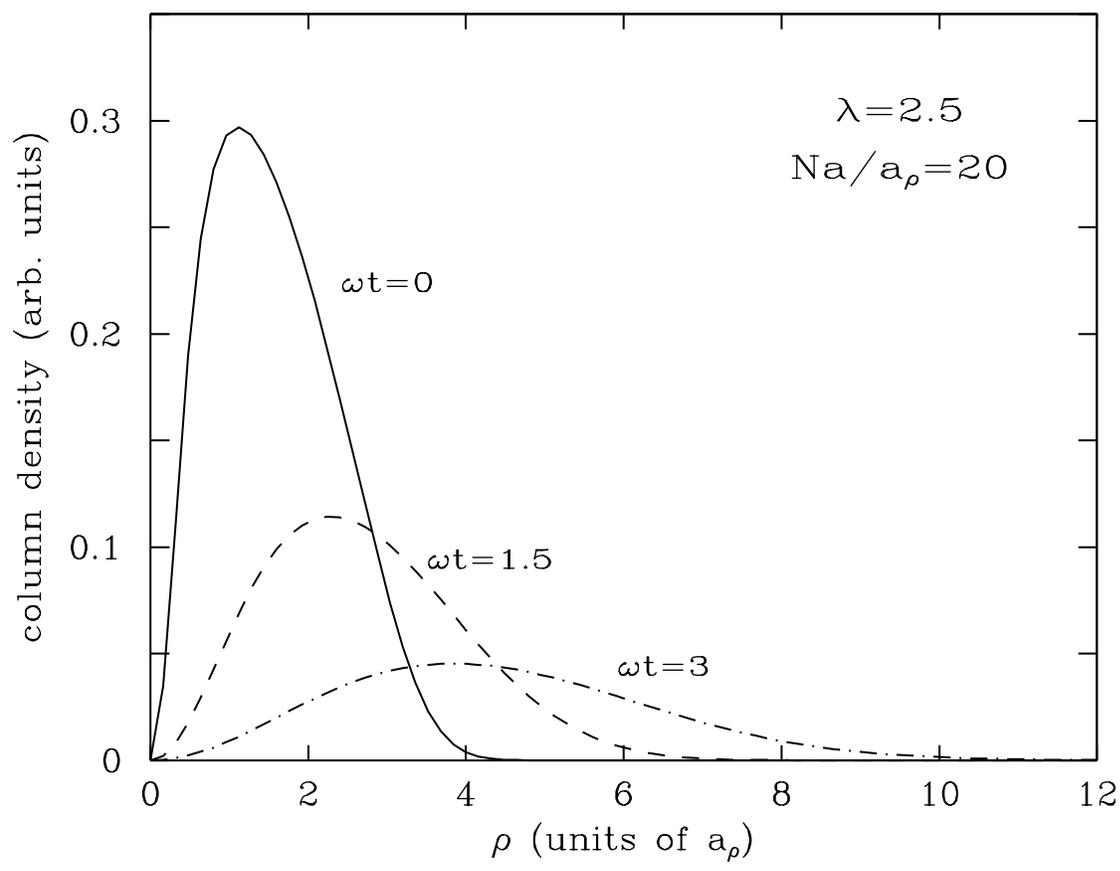
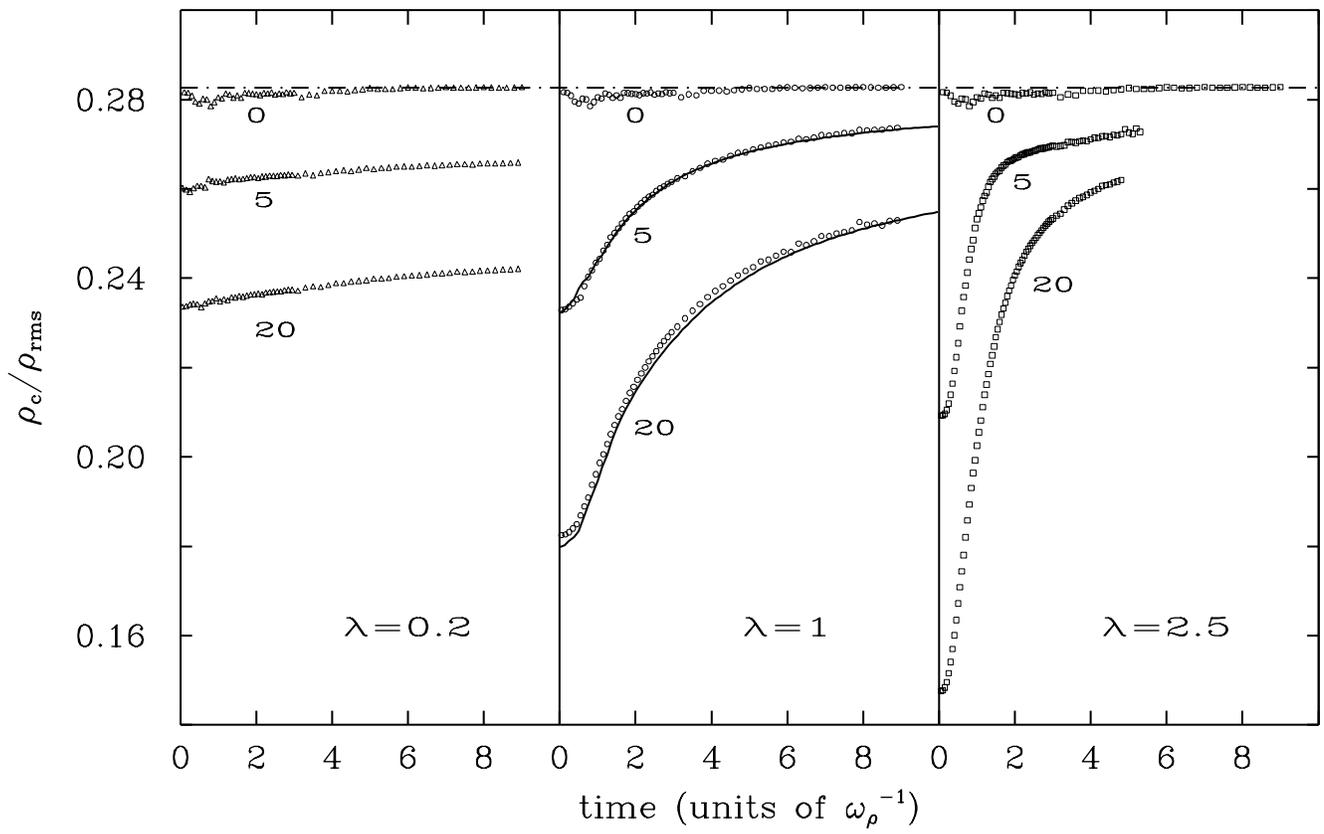
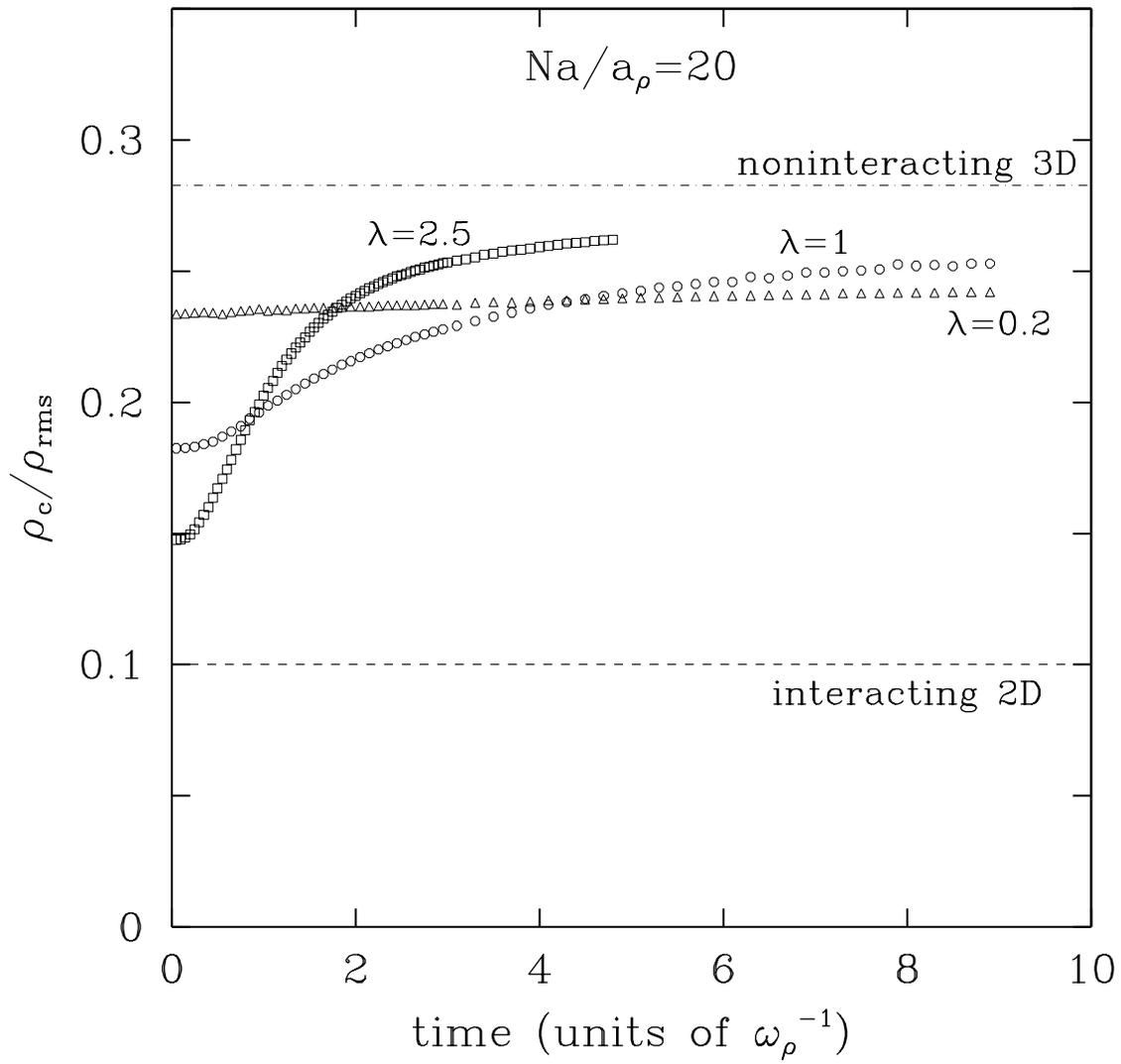
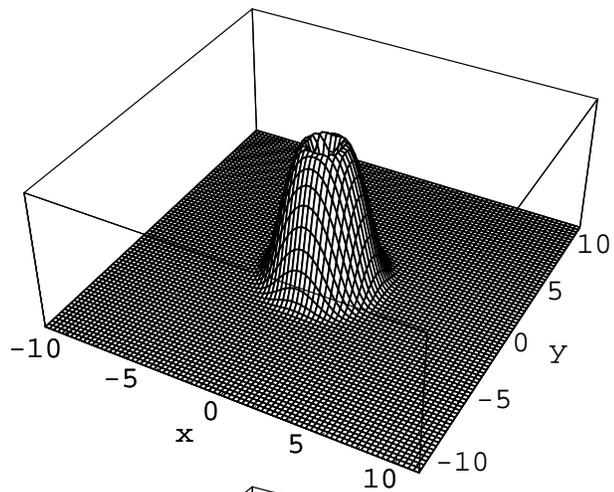
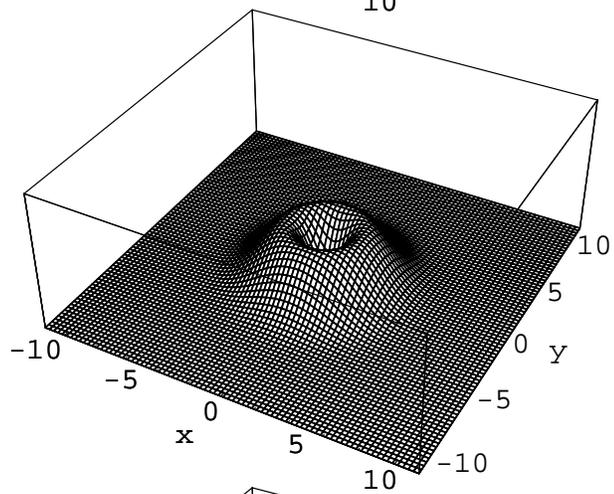
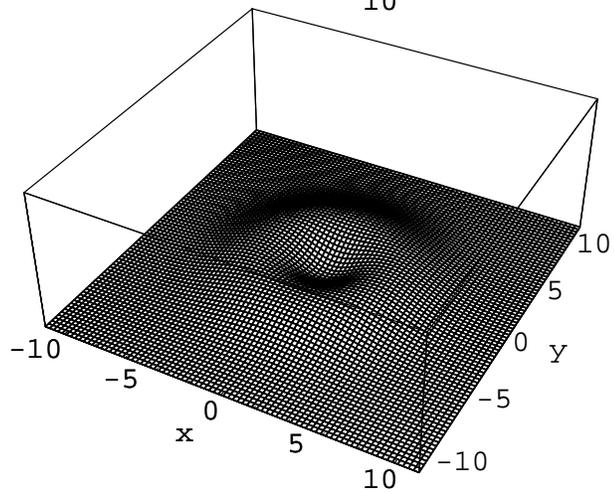
\end{document}